\documentclass[a4paper,11pt]{article}
\usepackage{pos}
\usepackage{graphicx}
\usepackage{subcaption}
\usepackage{tikz}

\let\OLDthebibliography\thebibliography
\renewcommand\thebibliography[1]{
  \OLDthebibliography{#1}
  \setlength{\parskip}{0pt}
  \setlength{\itemsep}{-0.7pt plus 0.3ex}
}

\title{Parametrising CCQE uncertainties in the Spectral Function model for neutrino oscillation analyses}
\ShortTitle{Parametrising CCQE uncertainties in the SF model for neutrino oscillation analyses}

\author*[a]{Jaafar Chakrani}
\author[a]{Margherita Buizza Avanzini}
\author[b]{Stephen Dolan}

\affiliation[a]{Laboratoire Leprince-Ringuet, IN2P3-CNRS, Ecole polytechnique, Institut Polytechnique de Paris,\\
Route de Saclay, 91128 Palaiseau Cedex, France}

\affiliation[b]{European Organization for Nuclear Research (CERN),\\
1211 Geneva 23, Switzerland}

\emailAdd{jaafar.chakrani@polytechnique.edu}

\abstract{
A substantial fraction of systematic uncertainties in neutrino oscillation experiments stems from the lack of precision in modeling the nucleus when describing the neutrino-nucleus interactions. The Spectral Function (SF) model features a distribution of momenta and removal energies of nucleons inside the nucleus within the shell-model picture, and also accounts for short-range correlations between nucleons. These characteristics offer significant improvements with respect to the more commonly used Fermi gas-based models. Electron scattering experiments offer a precise probe of the structure of the nucleus and have been used to both construct and validate the SF model. SF is thus an interesting reference model for long baseline neutrino experiments.

Based on constraints from electron scattering data, we develop a set of parameters that can alter the occupancy of the nuclear shells and the distribution of the nucleon momentum within each shell. In addition, the impact of final-state interactions on the outgoing lepton and nucleon kinematics, the contribution of short-range correlations and the effect of Pauli blocking can also be modified. In this document, we will first describe the development of these parameters, partially based on a comparison with electron scattering data. We then show fits of these parameters to available T2K and MINER$\nu$A cross-section data and discuss how they can be used to constrain the systematic uncertainties related to the SF model in neutrino oscillation analyses.
}

\FullConference{%
  *** The 22nd International Workshop on Neutrinos from Accelerators (NuFact2021) ***\\
  *** 6–11 Sep 2021 ***\\
  *** Cagliari, Italy ***}

\begin{document}
\maketitle

\section{Introduction}
Modeling neutrino-nucleus ($\nu$-A) interactions remains a major challenge for current and future neutrino oscillation experiments. The statistical uncertainties are still dominant in ongoing experiments, but the expected statistics in the next-generation experiments, like Hyper-Kamiokande \cite{hkref} and DUNE \cite{duneref}, show that the uncertainties on the neutrino interactions will be the limiting factor hindering a precise measurement of the oscillation parameters such as the CP-violating phase. To tackle this issue, experiments are currently constructing or designing more sophisticated detectors, but more accurate models in neutrino event generators are also necessary to precisely take into account the various processes occurring when neutrinos interact with matter. 

At $\sim 1$ GeV, neutrinos interact with nucleons bound within nuclear targets (Carbon, Oxygen,~...) and nuclear effects have a significant impact on the kinematics of the final state particles; this affects the reconstruction of the incoming neutrino energy, which directly biases the measurement of neutrino oscillation parameters. In particular, due to the removal energy and the Fermi motion of the bound nucleon, the neutrino interacts on a non-static particle and this can affect the whole interaction kinematics. A sophisticated way to model the charged-current quasi-elastic interaction (CCQE) is the Benhar Spectral Function (SF) model \cite{Benhar:1994hw} which is built from electron scattering data and allows a detailed description of the initial nuclear state. Multiple effects can further affect the final state kinematics: this is the case of Pauli blocking (PB), which prevents some interactions to occur, and of final-state interactions (FSI), which describe how the interaction between the struck nucleon and residual nucleus changes the cross section. 

In these proceedings, we introduce a parameterisation of systematic uncertainties for the SF model that can impact observables such as the lepton kinematics and the single transverse variables. 
We then show how the use of this parameterisation in fits to data can improve the agreement of the SF model with recent measurements from T2K and MINER$\nu$A experiments. Indeed, with such freedoms, uncertainties on the input models can be provided for future neutrino oscillation analyses. 


\section{Nuclear effects and systematic uncertainties}
In this work, the $\nu$-A interaction cross sections and the kinematics of the outgoing particles are simulated using \texttt{NEUT} \texttt{5.4.0} neutrino event generator. It describes CCQE $\nu$-A interactions with the SF model with an axial mass $M_A^{QE}$ for the quasi-elastic processes set at 1.03~GeV and a prior uncertainty of $\pm 0.06$~GeV, estimated from bubble chamber data.

The SF model provides a description for the two-dimensional distribution of removal energy ($E_m$) and momentum ($p_m$) of the nucleons within the Carbon (C) and Oxygen (O) nuclei. This accounts for the Fermi motion of the initial state nucleon as well as the removal energy needed to liberate the target nucleons from the nucleus which is distributed into nuclear shells; within each shell, the nucleon momentum (or missing momentum) distribution can be different. The SF model also accounts for short-range correlations (SRC) where the target nucleon is bounded within a strongly-correlated pair resulting in two nucleons in the final state. 

Since the SF model is built from electron scattering $(e, e'p)$ data, inputs from such experiments are used to evaluate the model uncertainties. First, the uncertainty on the shell occupancies is parameterised by a normalisation factor for each shell content. The prior uncertainties are then chosen to cover the shape differences of the measured $E_m$ distributions reported in Ref.\ \cite{Dutta:2003yt}. Another parameter is introduced to alter the SRC event normalisation with a wide $50\%$ prior and accounts for the discrepancies with other event generator predictions. 
$(e,e'p)$ data from Ref.\ \cite{Dutta:2003yt} are also used to assess the uncertainty on the $p_m$ distribution.
By choosing two ``extreme'' distributions of the measured missing momentum in each shell, a parameter is defined to allow a change in the shape of the $p_m$ distribution within these boundaries.

\texttt{NEUT} features a simple description of PB, inspired by the Relativistic Fermi Gas model of the nuclear ground state, where the cross section for portion of the phase space in which the pre-FSI outgoing primary nucleon has a lower momentum than Fermi momentum $p_F$ (209 MeV/c for C and O) is set to 0. Within this simplistic picture, a parameter varying the threshold $p_F$ is defined with a conservative prior of $\pm30$ MeV/c for each target.

FSI play a crucial role in altering the hadron kinematics and distorting interaction topologies and are modeled in \texttt{NEUT} with an intranuclear cascade. To simply account for FSI uncertainty, events are divided into two classes: ``{With FSI}'' and ``{Without FSI}''. A normalisation parameter for each class is introduced.  These two parameters have a broad $30\%$ uncertainty and are also correlated to ensure that the total cross section remains constant per each interaction mode. 
This has a similar impact as changing the mean free path of the nucleon within the nucleus.

The SF model is based on the plane-wave impulse approximation (PWIA) where the impact of FSI is not included in the cross section predictions and is only considered with an intranuclear cascade model. This distorts the outgoing nucleon momentum distribution and accounts for additional hadron ejection but it does not allow the inclusive CCQE cross section to vary as a full treatment of the distortion of the outgoing nucleon wave-function would. To account for this missing alteration to the cross section and for physics beyond the PWIA, an approach has been developed in Ref.\ \cite{Ankowski:2014yfa} to modify the SF prediction on the lepton kinematics via an optical potential. The parameterisation here corresponds to the strength by which this correction is applied to the \texttt{NEUT} model.


\section{Fits to published cross section measurements}

The SF parameterisation described above is used to fit the O and C cross section data, as measured by T2K in muon momentum ($p_\mu$) and direction ($\cos \theta_\mu$) \cite{t2k:jointoc}, as well as C cross section data in the transverse momentum imbalance ($\delta p_T$) from both T2K \cite{t2k:dpt} and MINER$\nu$A \cite{mnv:dpt}.
In these measurements, the topology of interest is the charged-current without pion in the final state (CC0$\pi$), which also contains events from 2p2h and charged-current resonant (CCRES) processes. 
Additional uncertainties related to those channels are thus included in the fit via normalisation parameters for 2p2h as well as CCRES events with a pion absorbed through FSI.

\texttt{NUISANCE} \cite{nuisance} is used to process \texttt{NEUT} outputs, define this parameterisation of the uncertainties and perform fits to the recent cross section measurements from T2K and MINER$\nu$A. 

\subsection{Chi-squared test-statistic}

 
In order to avoid Peelle's Pertinient Puzzle \cite{ppp} which artificially favours low normalisation fit results, the covariance matrix as released by the experiments is (non-linearly) transformed into a matrix with a normalization component, a shape block, and the correlation between the two. This results in a relative uncertainty that remains constant as a function of the normalisation, as motivated by the arguments of Ref.\ \cite{ppp}, thus mitigating the problem (although with caveats beyond the scope of these proceedings).
To do so, if $B=(B_1, ...,B_n)$ are the measured $n$ bins and $M$ their covariance matrix,  we define a vector corresponding to the relative normalization of each of the first $n-1$ bins and the total normalization in the last bin: $C = \left(B_1/B_T, ..., B_{n-1}/B_T, B_T\right)$ where $B_T = \sum_{i=1}^n B_i$. The covariance matrix $N$ of the vector $C$ can be computed within the linear approximation using the Jacobian of the function $B \mapsto C$:
$N_{ij} = \sum_{k,l} \frac{\partial C_{i}}{\partial B_k} M_{ij} \frac{\partial C_{j}}{\partial B_l}. $
The matrix $N$ has the same dimension and the same positive-definiteness properties as $M$ since the mapping $B \mapsto C$ is a bijection. 

In this work, the fits optimise the chi-squared test statistic in these transformed coordinates. It is defined as $\chi^2_\text{NS} = \chi^2_\text{data, NS} + \chi_\text{syst}^2$ where 
$\chi^2_\text{data, NS} = \sum_{i,j} \left(C_i - C_i^\text{MC}\right) \left( N^{-1} \right)_{ij} \left(C_j - C_j^\text{MC}\right)$ and $\chi^2_\text{syst}$ encodes the prior knowledge on the fit parameters including the Gaussian uncertainties.

\subsection{Fit results}

\begin{figure}[!t]
    \centering
    \begin{subfigure}[b]{0.22\textwidth}
        \begin{tikzpicture}
            \draw (0,0) node[inner sep=0] {\includegraphics[height=4cm, trim={1.6cm 0 0 0}, clip=false]{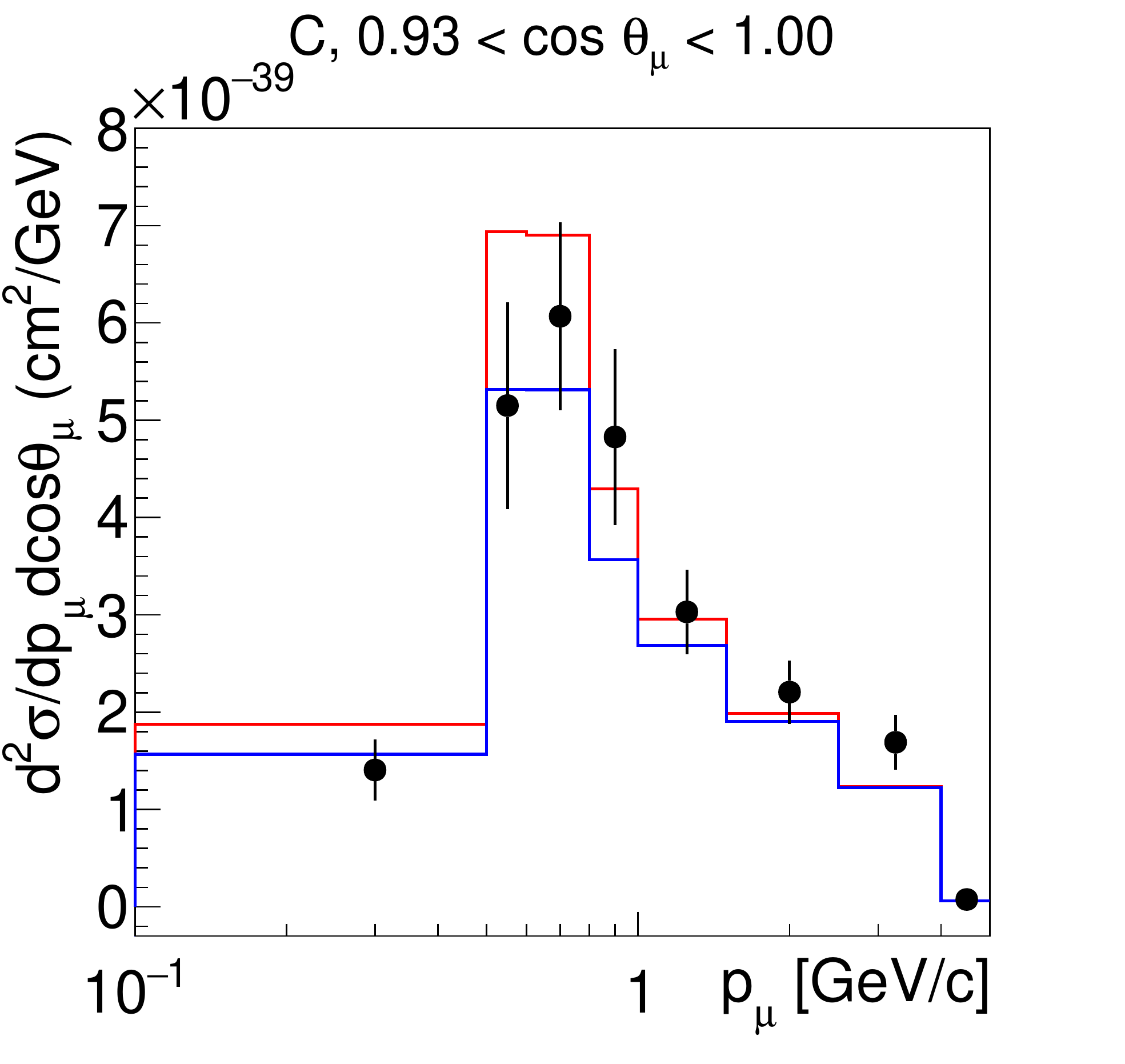}};
            \draw (-0.4cm, -2.15cm) node {(a)};
        \end{tikzpicture}
    \end{subfigure}
    \begin{subfigure}[b]{0.23\textwidth}
        \begin{tikzpicture}
            \draw (0,0) node[inner sep=0] {\includegraphics[height=4cm, trim={1.55cm 0 0 0}, clip]{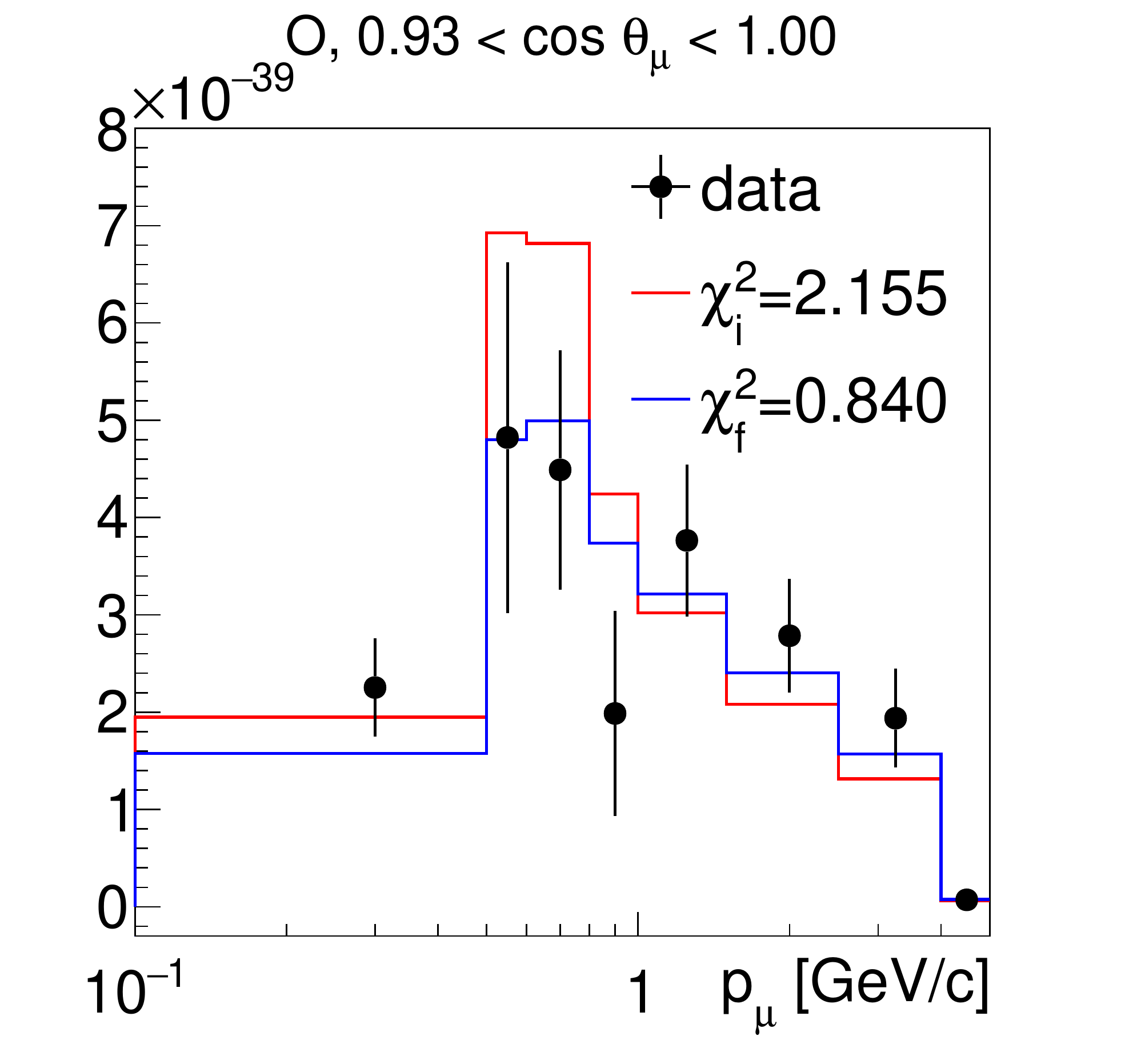}};
            \draw (-0.4cm, -2.15cm) node {(b)};
        \end{tikzpicture}
    \end{subfigure}
    \begin{subfigure}[b]{0.28\textwidth}
        \begin{tikzpicture}
            \draw (0,0) node[inner sep=0] {\includegraphics[height=4cm, trim={0 0.5cm 0 1.2cm}, clip]{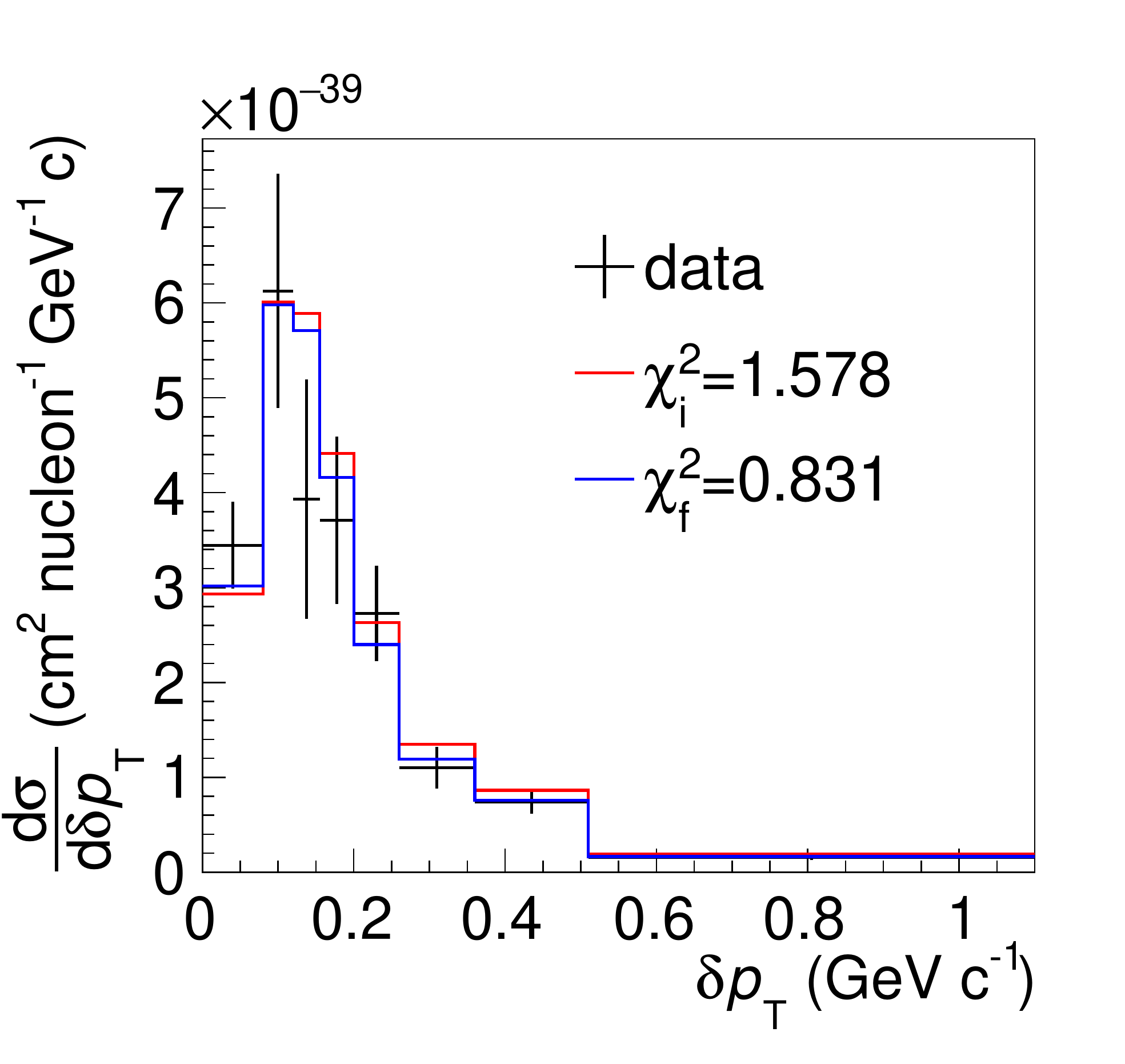}};
            \draw (0.4cm, -2.15cm) node {(c)};
        \end{tikzpicture}
    \end{subfigure}
        \begin{subfigure}[b]{0.25\textwidth}
        \begin{tikzpicture}
            \draw (0,0) node[inner sep=0] {\includegraphics[height=4cm, trim={2.8cm 0.5cm 0 1.2cm}, clip]{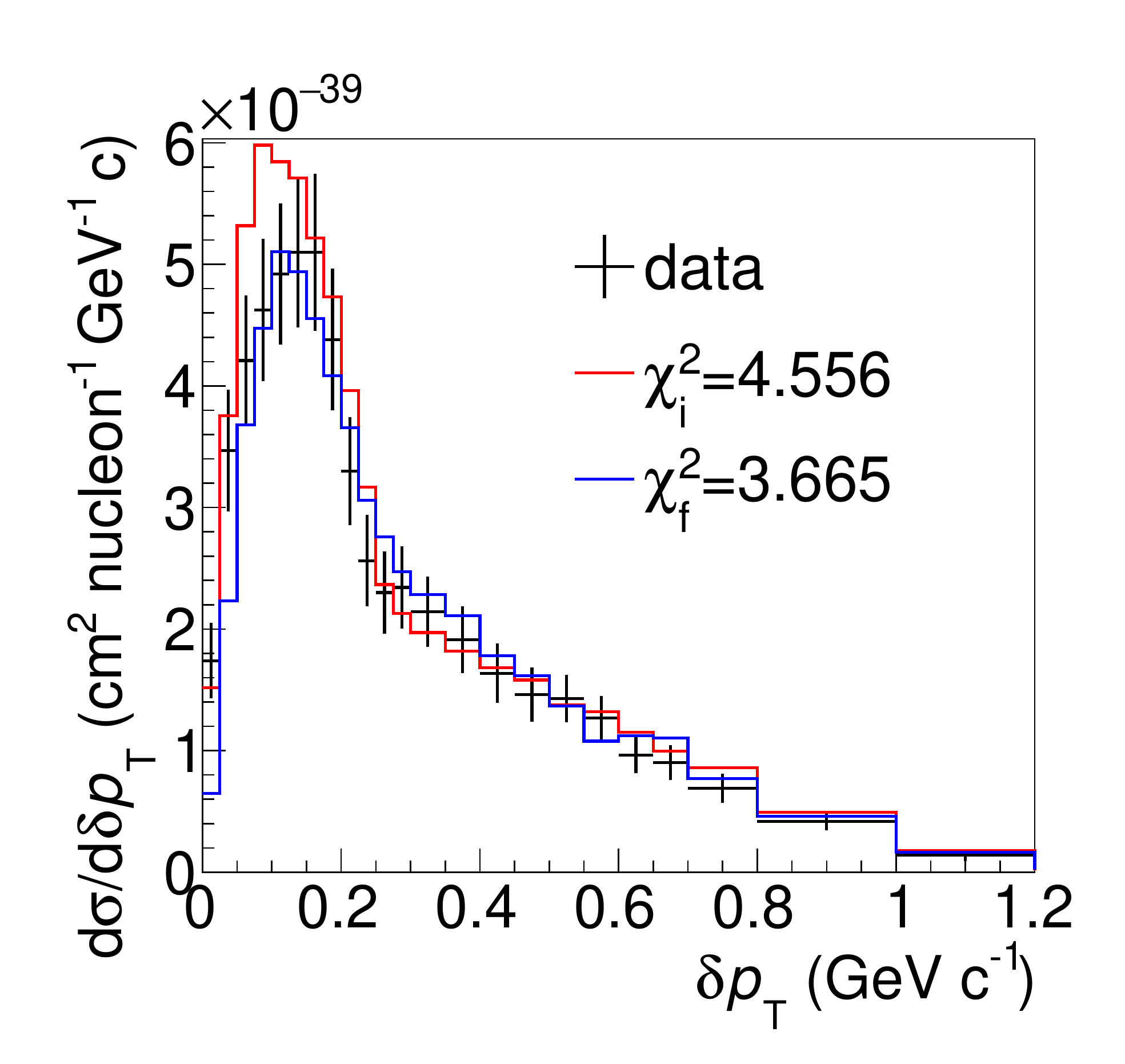}};
            \draw (-0.3cm, -2.15cm) node {(d)};
        \end{tikzpicture}
    \end{subfigure}
    \caption{Prefit (red) vs.\ postfit (blue) distributions for the forward bins in the T2K joint O-C measurement (a \& b) and for $\delta p_T$ data by T2K (c) and MINER$\nu$A (d). The quoted chi-squared $\chi^2_i$ (resp.\ $\chi^2_f$) corresponds to the prefit (resp.\ postfit) $\chi^2_\text{data, NS}$ divided by the number of degrees of freedom.}
    \label{fig:prepostdists}
\end{figure}
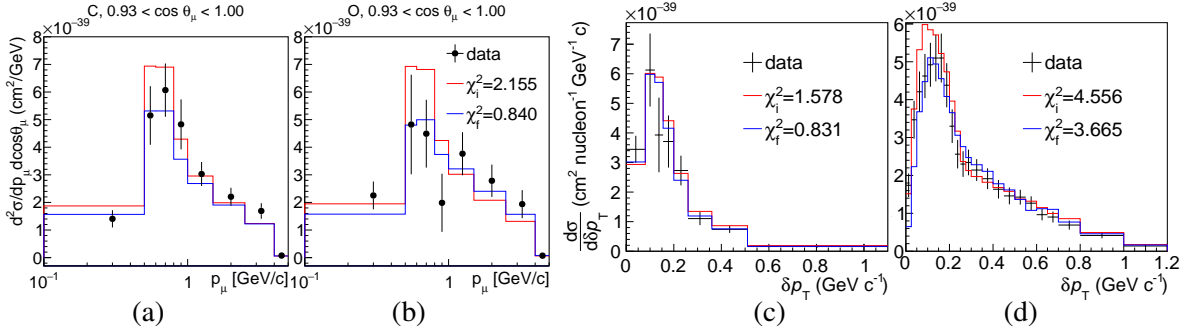

\begin{figure}[ht]
    \centering
    \begin{subfigure}[b]{0.64\textwidth}
        \includegraphics[height=6.5cm, trim={1.6cm 0 2.1cm 0.5cm}, clip]{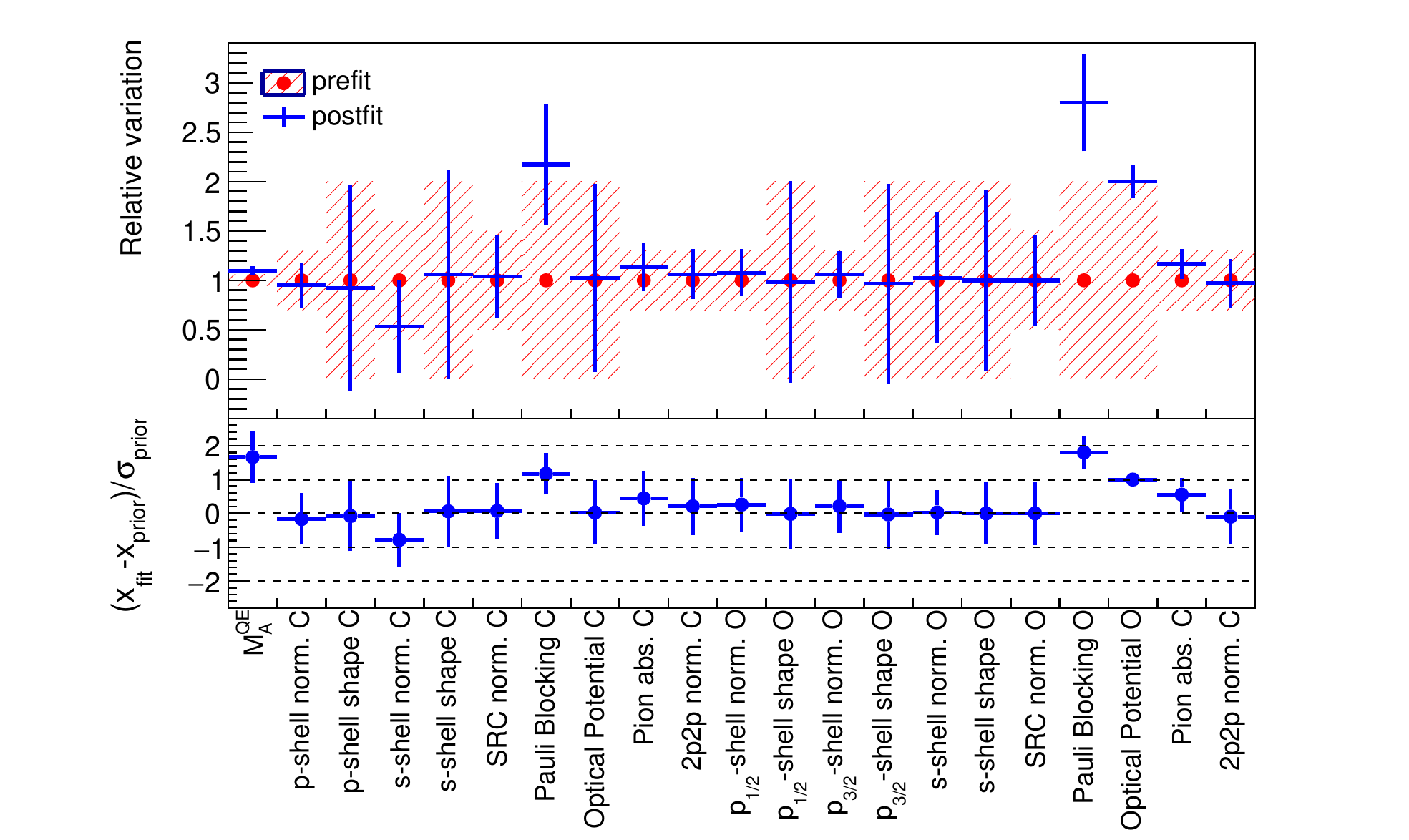}
    \end{subfigure}
    \begin{subfigure}[b]{0.35\textwidth}
        \includegraphics[height=6.5cm, trim={2.8cm 0 2cm 0.5cm}, clip]{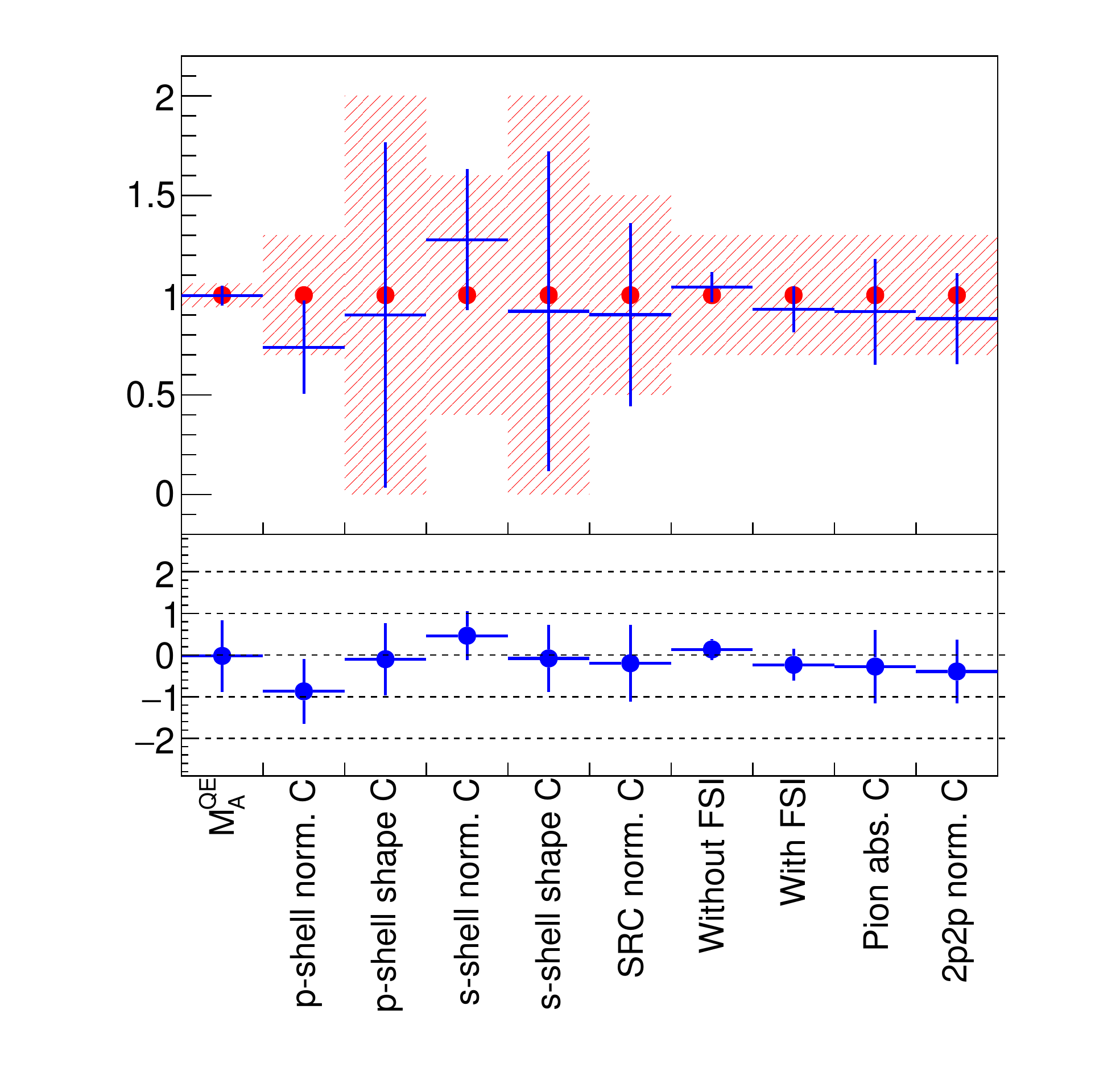}
    \end{subfigure}
    \caption{Prefit vs.\ postfit values and constraints from T2K joint O-C (left) and T2K $\delta p_T$ (right) data. }
    \label{fig:prepostparams}
\end{figure}

In Fig.\ \ref{fig:prepostdists} we report the prefit and postfit cross section distributions for the considered data. In particular, for the joint O-C measurement, only the highest $\cos \theta_\mu$ bins are shown (a \& b). This forward region corresponds to low momentum exchange ($Q^2$) where models struggle the most in reproducing data. This is confirmed by the high contribution of this bin in the prefit chi-squared. The postfit chi-squared indicates a significantly better data/model agreement and, as shown in Fig.~\ref{fig:prepostparams} (left), this improvement is driven by the important change of the Pauli blocking and optical potential parameters, which are particularly affecting the low $Q^2$ region.
A fair sensitivity to the shell normalisation dials is also visible in Fig.\ \ref{fig:prepostparams}. On the other hand, there is no noticeable sensitivity to the missing momentum shape because the constraints from electron scattering data are tight and because the muon kinematics are not notably sensitive to the shape of the $p_m$ distribution. 

Measuring $\delta p_T$ probes the nuclear effects undergone by the knocked-out nucleons. It gives access to the transverse component of the initial state nucleon momentum (Fermi motion) which contributes to the bulk part of $\delta p_T$ distribution. It also gives information on SRC, 2p2h and FSI processes in the tail part of its distribution. 
As shown in Fig.\ \ref{fig:prepostdists} (c) for T2K, the prefit chi-squared value is already low, but the data/model agreement is further improved after the fit. The parameters used in this fit and their prior and postfit constraints are shown in Fig.\ \ref{fig:prepostparams} (right), where we see that the sensitivity to the different parameters is limited by the low statistics of the measurement.

The sensitivity improves when using MINER$\nu$A data which has larger statistics. However the postfit data/model agreement remains limited (see Fig.\ \ref{fig:prepostdists} (d)). This can be attributed to the fact that, due to the higher energy of the neutrino flux in MINER$\nu$A, there is a higher contribution in the CC0$\pi$ sample from CCRES interactions undergoing pion absorption in comparison with T2K. This can also indicate that the presented parameterisation of the CCQE model may need further improvements, especially the FSI effects.

\section{Conclusion}
We introduced a parameterisation of systematic uncertainties on the inclusive and semi-inclusive predictions of the SF model. They allow an improved agreement with data after fitting to T2K cross section measurements. Indeed, the described parameterisation has been adopted by the T2K collaboration and will be used for the next oscillation analysis. However, MINER$\nu$A data suggests that this description is still incomplete and may need further improvements particularly on the nucleon FSI processes. Nevertheless, the uncertainty model presented here represents a robust starting point for CCQE uncertainties in neutrino oscillation analyses.


\begin{thebibliography}{99}

\bibitem{hkref}
Abe, K. et al., arXiv:1805.04163.

\bibitem{duneref}
Abi, B. et al., arXiv:2002.03005.

\bibitem{Benhar:1994hw}
Benhar, O. et al., Nucl. Phys., A579:493–517, 1994.

\bibitem{Dutta:2003yt}
Dutta, D. et al., Phys. Rev., C68:064603, 2003. 

\bibitem{Ankowski:2014yfa}
Ankowski, A. et al., Phys. Rev., D91(3):033005, 2015.

\bibitem{t2k:jointoc}
Abe, K. et al., Phys. Rev., D101:112004, 2020.

\bibitem{t2k:dpt}
Abe, K. et al., Phys. Rev., D98:032003, 2018.

\bibitem{mnv:dpt}
Lu, X. G. et al., Phys. Rev. Lett., 121:022504, 2018.

\bibitem{nuisance}
Stowell, P. et al., JINST, 12 P01016, 2017.

\bibitem{ppp} 
D’Agostini, G., Nucl. Instrum. Meth., A 346, p306, 1994.

\end{thebibliography}
\end{document}